\documentclass[11pt]{article}
\usepackage{amsmath}
\usepackage{multicol}
\usepackage{graphicx}
\usepackage{pstricks}
\usepackage{pst-plot}
\usepackage{floatflt}
\usepackage{epsfig}
\usepackage[french]{babel}
\usepackage[latin1]{inputenc}

\def\bq{\begin{quote}\em\small}
\def\eq{\end{quote}}

\def\be{\begin{equation}}
\def\ee{\end{equation}}

\def\bea{\begin{eqnarray}}
\def\eea{\end{eqnarray}}
\def\beann{\begin{eqnarray*}}
\def\eeann{\end{eqnarray*}}

\newcommand{\ket}[1]{|\kern.3ex #1 \kern.3ex\rangle}
\newcommand{\bra}[1]{\langle\kern.3ex #1 \kern.3ex|}
\newcommand{\braket}[2]{\langle\kern.3ex #1 \kern.3ex|
                        \kern.3ex #2 \kern.3ex\rangle}

\begin{document}

\title{
EINSTEIN ET LA MÉCANIQUE QUANTIQUE.\\ Quelques repères, quelques
surprises.\\ {\small Séminaire donné le 8 septembre 2005 dans le cadre de
l'Année de la Physique.} }
\author{ P.Roussel,
Institut de Physique nucl\'eaire,
Universit\'e Paris XI, CNRS, IN2P3 \\
\hspace{5.5cm}F-91406 Orsay Cedex\hspace{2.5cm} {\bf IPNO-DR-06-01}
}
\date{}

\vspace{-2cm}

\maketitle
\begin {abstract}
\hspace{-2.5cm}\begin{minipage}{17.1cm}
\small Si Einstein a apporté des contributions majeures et même fondatrices à la
mécanique quantique, il est aussi resté profondément insatisfait des bases
de cet édifice qu'il savait pourtant être si efficace pour la physique. Ses
critiques sont surtout connues au travers des nombreuses controverses qu'il
a poursuivies avec N. Bohr pendant plusieurs décennies.

On a laissé entendre que ses critiques étaient celles d'un vieil homme
dépassé et que de toute façon cela relevait de la philosophie. Et puis,
grâce à J. Bell, on a dit que le débat relevait maintenant de l'expérience,
curieuse intrusion expérimentale dans la philosophie, ou curieuse
philosophie et que enfin, l'expérience réalisée, notamment à Orsay avec les
succès répétés de A. Aspect, avait tranché et que c'était en faveur de Bohr.
Le débat serait-il clos?

Aujourd'hui, cinquante ans après sa mort, le regard qu'on porte sur les
critiques d'Einstein à la mécanique quantique semble en train de changer.
Pourquoi? Assiste-t-on au retour de la physique?

On tentera de préciser ce qu'a vraiment dit Einstein, et on tentera de
donner des éléments de réponse aux questions soulevées.
 
\vspace{.4cm}
{\em
\begin {center}
EINSTEIN and QUANTUM MECHANICS 
\end {center}
{\bf \em Abstract}\hspace{5cm}
Some landmarks, some surprises

\small Even if Einstein brought major contributions as a founder of quantum
mechanics, he remained deeply unsatisfied with the bases of this structure
he knew to be so efficient for physics. His critics are often known through
his numerous controversies with N. Bohr for a few decades.

It has long been suggested that his critics were obsolete and came from an
aging man and that it was more a question of philosophy anyway. After J.
Bell's contribution however it was claimed that the debate could be closed
by an experimental proof, a strange experimental intrusion in philosophy or
a strange philosophy. Finally, the experiment being performed, in particular
in Orsay with the repeatedly successful experiments of A. Aspect, it was
claimed that N. Bohr won. Would the debate be closed?

Today and fifty years after his death, it seems that Einstein's critics to
quantum mechanics are being regarded differently. Why is it so? Is it due to 
physics coming back to the debate?

It will be attempted to accurately report Einstein's writings on the matter
and to find  elements of answer to the raised questions.}

\end{minipage}
\end {abstract}

\newpage

\section {Introduction.}


``Einstein et la mécanique quantique''
...Pourquoi ce sujet à l'occasion de l'année de la physique. C'est qu'il
donne l'exemple d'une physique vivante mettant en rapport des gens
entre eux et pas seulement leur rapport aux choses, et même, les
rapports organisés des gens dans ce qu'on peut appeler une
Institution. Avec ce sujet, on peut aussi montrer, à côté de ses
succès, les difficultés, les doutes, les confusions de la physique en
train de se faire et de la société des physiciens qui produit cette
physique. La physique est le résultat d'une activité humaine. Le sujet
est vaste, il a fallu faire des choix difficiles et en partie
arbitraires.

Dans une interview, il y a quelques mois, Françoise Balibar qui a passé
beaucoup de sa vie à travailler sur Einstein et, pourrait-on dire, pour
Einstein (elle est un des éditeurs des oeuvres complètes d'Einstein en
français), notait que le regard sur Einstein et ses objections changeait et
qu'on ne les considérait plus tellement comme
\bq "celles d'un vieil homme décalé par rapport à son temps".\eq
Mais si alors quelque chose a changé, cinquante ans après sa mort, ce n'est
en tous cas pas lui, Einstein.

Dans cette introduction, je voudrais montrer par quelques citations, qu'en
effet, on ne dit plus la même chose sur les rapports d'Einstein à la
mécanique quantique et puis par d'autres citations, évoquer quelques unes des
personnalités qui, avec Einstein, ont été des acteurs de cette période
historique de la physique.

Werner Heisenberg (un cher collègue, vous allez voir!) en 1969 :

\bq ``Tout travail scientifique repose, consciemment ou non, sur une option
philosophique, une certaine structure mentale, qui fournit un appui ferme à
cette pensée. Il peut arriver cependant, à mesure que la science progresse,
qu'un nouveau domaine d'investigation ne soit parfaitement compréhensible
que si le chercheur fournit l'effort énorme d'élargir ce cadre et de
modifier la structure même de sa pensée. Il semble qu'Einstein, dans le cas
de la mécanique quantique, n'était plus disposé, ou plus apte, à franchir ce
pas'' {\em \cite{heis1969}}.\eq

"plus disposé ou plus apte" on n'est effectivement pas loin du "vieil
homme dépassé"!

Mais qu'est ce qui lui est reproché? On trouve la réponse quelques
lignes plus haut (c'est donc toujours W. Heisenberg en 1969) :

\bq ``...il [Einstein] refusait de voir dans la mécanique quantique une
représentation {\bf définitive et exhaustive} des phénomènes physiques. Ses
conceptions philosophiques impliquaient, d'une part, la conviction
qu'il est possible de diviser radicalement le monde en {\bf un domaine
objectif et un domaine subjectif}, d'autre part, l'hypothèse que l'on
doit pouvoir décrire l'aspect objectif de façon non équivoque. La
mécanique quantique ne pouvait satisfaire à cette double exigence, il
ne semble pas que la science puisse jamais retrouver le chemin des
postulats d'Einstein'' {\em \cite{heis1969}}.\eq   

Voyez que ce n'est pas tant de la physique qui lui est reprochée que
de la philosophie, mais il est vrai, la physique n'est pas loin : si la
distinction objectif/subjectif doit être révisée, faut-il attribuer un
rôle à la conscience dans la physique quantique? Et puis la physique
doit-elle traiter, au moins en principe, les cas individuels ou bien
peut-elle se contenter des ensembles statistiques?

 Mais n'êtes vous pas un peu choqués de cette idée (en 1969) d'une
représentation "définitive et exhaustive"? Elle vient de loin
pourtant, déjà en 1927 (congrès Solvay), le même Heisenberg avec
M.Born :

\bq ``Nous tenons la mécanique des quanta pour une théorie complète dont
les hypothèses fondamentales physiques et mathématiques ne sont plus
susceptibles de modifications'' {\em \cite{paty1985} }.\eq

Mais N. Bohr n'est-il pas dans le même état d'esprit en 1935 dans sa          
réponse à Einstein, Podolsky, Rosen (on y reviendra!) quand il écrit :

\bq ``... l'argumentation d' EPR pourrait difficilement prétendre affecter
la solidité de la description fournie par la mécanique quantique qui est
fondée sur un formalisme mathématique cohérent couvrant automatiquement
toute procédure de mesure y compris celle indiquée par EPR'' {\em
\cite{bohr1935}}.\eq

Avec Einstein, le ton est tout différent.
En 1926, dans une lettre à M. Born :

\bq ``La mécanique quantique force le respect. Mais une voix intérieure me
dit que ce n'est pas encore le nec plus ultra. La théorie nous apporte
beaucoup de chose, mais elle nous rapproche à peine du secret du
Vieux. De toute façon, je suis convaincu que lui, au moins, ne joue
pas au dés''  {\em \cite{paty1985} page 335}.\eq 

On a souvent cité cette dernière phrase, on y reviendra plus loin.
En 1933 dans une lettre où Einstein refuse la proposition qui lui est faite,
à l'initiative de Paul Langevin, de venir s'installer en France pour
échapper aux persécutions nazies (il ira finalement à Princeton).

\bq ``J'ai beaucoup travaillé, certes, mais j'ai aussi rejeté la plus
grande partie de ce que j'ai fait ; et je ne sais nullement encore si
l'avenir me donnera raison sur ce qui reste. En fait, je ne possède
ni une compétence ni un savoir particulier, mais {\em seulement} la passion
de la recherche'' {\em \cite{merl1993}}.\eq

En 1935 enfin, le titre de ce fameux article EPR ne reste-t-il pas
sous forme d'interrogation : \bq "La description fournie par la mécanique
quantique de la réalité physique peut-elle être considérée comme
complète?'' {\em \cite{EPR} (Bohr il est vrai y répondra avec le même
titre!).}\eq

On aurait tort cependant de ne voir dans ces différences que le reflet de
personnalités différentes (et des circonstances quelquefois!). C'est bien
aussi l'effet de l'Institution qui d'un côté repousse et refuse le nouveau
tant qu'elle le peut (elle n'aime pas non plus le doute, on en verra un
exemple tout à l'heure), et puis quand elle l'accepte, elle le rigidifie et
le dogmatise (le ``paradigme'' du contesté Thomas Kuhn!).

Mais revenons au "vieil homme dépassé", est-ce le ton aujourd'hui?  On
trouve bien encore des titres accrocheurs qui pourraient rappeler cet
état d'esprit : 

\bq ``EINSTEIN N'ACCEPTE PAS LA MÉCANIQUE QUANTIQUE'' {\em \cite{pls2002}} \eq

mais quand on parle vraiment de physique, ici, ce sont Alain Aspect et
Philippe Grangier qui s'expriment, c'est tout différent.
Le titre déjà : \bq ``DES INTUITIONS D'EINSTEIN À L'INFORMATION QUANTIQUE :
LES STUPÉFIANTES PROPRIÉTÉS DE L'INTRICATION'' {\em \cite{bup2005}}.\eq

La première phrase du résumé : \bq ``En 1935, dans un article célèbre,
EINSTEIN, PODOLSKY et ROSEN {\bf introduisent la notion d'état intriqué},
dans lequel deux particules présentent des corrélations fortes même
si elles sont très éloignées''.\eq

Plus loin dans le texte, on indique que ``l'état intriqué" est tout
aussi bien appelé aujourd'hui "état EPR".
On est loin du vieil homme dépassé!
Enfin, sur la mécanique quantique déjà définitive en 1969 ou même en 1927 :

\bq ``L'interaction d'un appareil de mesure classique et d'un objet
quantique a des conséquences dramatiques sur la fonction d'onde de l'objet
mesuré, mais {\bf qui est loin d'être parfaitement comprise}''
{\em \cite{lcp2005}}.\eq
On est en 2005!

\section {Repères historiques.}

Einstein et la mécanique quantique. Il est utile de garder en tête quelques
repères historiques qui jalonnent la question sur un siècle. En 1905,
Einstein a 26 ans et contribue cette année là à la naissance de la mécanique
quantique (bien sûr, elle ne s'appelle pas encore comme cela!). Einstein
décède un demi-siècle plus tard en 1955. Quand en 2005 on regarde en
arrière, on voit un demi-siècle avec Einstein et un demi-siècle sans
Einstein. On va plus s'intéresser au premier -c'est le sujet qui veut cela-
mais en prenant ce qui se dit aujourd'hui comme référence chaque fois que
cela sera utile. On a fait figurer D. Bohm dans les deux demi-siècles, d'une
part parce qu'il a joué probablement un rôle charnière, et que d'autre part,
ce rôle est largement ignoré.

\subsection {Un demi-siècle avec Einstein.}

On pourrait garder l'impression que dans la première moitié de ce
demi-siècle Einstein a apporté successivement les éléments constitutifs de
la mécanique quantique et que dans la seconde il a surtout critiqué. On
verra tout à l'heure qu'il était critique et plein de doute dans la phase de
construction et conscient des succès dans la phase critique. Mais ici,
survolons d'abord les événements.
\begin {itemize}
\item {\bf 1905} Einstein pose, pourrait-on dire, la première pierre de la mécanique
quantique (elle ne s'appelle évidemment pas comme cela) avec l'affirmation
que l'énergie du rayonnement est réellement quantifiée, ce n'est pas qu'un
artifice de calcul.

\item {\bf 1909} Einstein démontre la double nature, corpusculaire et ondulatoire, du
rayonnement à partir de l'évaluation des fluctuations de l'énergie.

\item {\bf  1916} Einstein démontre que le rayonnement est une «vraie »  particule (pour
ce qui est de sa nature corpusculaire!) avec non seulement une énergie mais
aussi une impulsion.

\item {\bf 1917 }Einstein établit les deux mécanismes de décroissance d'un état atomique
excité : émission stimulée et émission spontanée. On ne reparlera pas du
tout de cette question dont on sait pourtant toute l'importance dans la
physique, la technologie et la vie quotidienne d'aujourd'hui avec les lasers.

\item {\bf 1921-1923}  Prix Nobel pour l'année 1921 mais remis en juillet 1923 :

"Pour les services qu'il a rendu à  la Physique théorique et en particulier
pour sa découverte de la loi de l'effet photoélectrique."

\item {\bf 1924-25} L'expérience de Bothe et Geiger donne raison à Einstein contre
Bohr, Kramers et Slater, on reviendra plus loin sur le contenu de ce
différent et les bonnes raisons qu'avait Einstein de triompher dans un débat
qui avait débordé du seul monde scientifique pour atteindre ce qu'on
appellerait aujourd'hui les médias!

\item {\bf 1924 }(voir plus loin en 1995!)  Condensation Einstein-Bose.

\item {\bf 1927}  5ème Congrès Solvay. Une des circonstances célèbres où les fondateurs
de la mécanique quantique et en particulier Einstein et Bohr s'affrontèrent.

\item {\bf 1935} L'article Einstein Podolsky Rosen (EPR), on en reparlera, qui introduit
un élément nouveau dans les débats et qui est à la source de la physique de
l"intrication".

\item {\bf 1946-1953} Entre 1946 et 1949, un nombre important de contributions
d'Einstein et de ses contemporains sont rassemblées à l'occasion de son
70ème anniversaire. D'autres écrits d'Einstein apparaissent jusqu'en 1953
et même plus tard.

\item {\bf 1955 }Décès d'Albert Einstein à Princeton.

\end{itemize}

$~$

\hspace{1.cm}\begin{minipage}{12cm}
	{\bf 1952-56} D. Bohm, J.P. Vigier et L. de Broglie établissent une
	mécanique quantique {\bf causale} : elle reproduit tous les
	résultats de la mécanique quantique sans aucun élément probabiliste,
	elle est déterministe, mais au prix de potentiels dont les
	effets se transmettent "instantanément" dans tout l'espace ; elle
	est "{\bf non-locale}".
\end{minipage}
\subsection {Un demi-siècle sans Einstein..}

\hspace{1.5cm} {\bf 1952-56} mécanique quantique causale et non-locale : D. Bohm, J.P.
	Vigier.

\hspace{1.5cm} {\bf 1955} Décès d'Albert Einstein.
\begin{itemize}
\item	{\bf 1952-1957} EPR devient EPRB puisque D. Bohm propose une version de
	l'expérience de pensée de EPR mettant en jeu des spins et des
	orientations de polarisation et qu'avec A. Aharonov ils apportent la
	première preuve expérimentale de "non-localité" EPR. Ils suggèrent
	de changer les paramètres de détection pendant le vol des particules.

\item	{\bf 1964} Théorème de J.S. Bell. A partir de l'expérience EPRB, Bell
	établit une relation générale, une inégalité, qui distingue la
	mécanique quantique de toute théorie à variables supplémentaires
	dans le passé commun des particules. Il montre que trois directions
	(au moins) de polariseurs sont nécessaires pour que cette
	distinction soit possible.

\item	{\bf 1976-83} Les expériences d'Alain Aspect et al. (pour ne citer que
	nos collègues français d'Orsay!) apportent des preuves de plus en
	plus contraignantes des effets EPR.

\item	{\bf 1985...} Expériences sur les atomes froids, les cavités, les champs à
	petits nombre de photons avec C Cohen-Tannoudji et al. C'est avec
	ces éléments que la physique de ``l'intrication" connaîtra de
	multiples développements ; mais aussi avec le point suivant :

\item	{\bf 1991-96} Le concept de dé-cohérence avec Zurek, Brune, Haroche,
	Raimond, Omnes pose en terme nouveaux les relations du microscopique
	et du macroscopique et, forcément, éclaire autrement le problème
	du statut de la fonction d'onde.

\item	{\bf 1995} Première expérience conduisant, avec des atomes alcalins
	refroidis, à l'observation d'une condensation Einstein-Bose
	prédite... en 1924! mais ceci est en dehors de notre sujet.

\item	{\bf 2000-2005} L'intrication conduit à la naissance de ``l'information
	quantique" et...
	...nous laisserons à méditer quelques écrits contemporains qui,
	peut-être, auraient provoqué quelques réflexions d'Einstein!.
\end{itemize}	
	
\section {Des débuts difficiles : d'une science hésitante à une science
dogmatique.}

On va montrer que si la science tend à devenir dogmatique, elle peut aussi
avoir des périodes d'hésitation et par exemple, que le premier quart de
siècle, avant 1930, n'est pas l'accumulation
régulière des éléments successifs qui constitueront l'édifice final comme on
a pu le laisser entendre tout à l'heure.

	En {\bf 1910}, dans une lettre à Laub, un jeune collègue, Einstein écrit le 3
	mars :

	\bq ``La théorie des quanta ne fait pas de doute pour
	moi'' {\em \cite{bali1989} note 3 page 114}.\eq 

C'est très bien, nous sommes cinq ans après la fondation! mais quelques mois
plus tard, le 4 novembre, s'adressant au même Laub, c'est tout différent :

	 \bq ``J'ai en ce moment de sérieux espoirs de résoudre le problème du
	 rayonnement, {\bf et cela sans les quanta de lumière}. Je suis
	 terriblement curieux de voir comment les choses vont se passer. Il
	 faudrait renoncer au principe de conservation de l'énergie sous sa
	 forme actuelle'' {\em \cite{bali1989} page 114}. \eq

	 Voyez que cinq ans après la fondation, le fondateur doute des
	 quanta, mais n'a pas peur par contre d'envisager de renoncer à la
	 conservation d'énergie "dans sa forme actuelle". C'est précisément
	 ce que proposeront Bohr, Kramers et Slater, dans un article cette
	 fois, en 1924. Einstein sera alors, on le sait, retourné aux quanta
	 et aura abandonné la révision des principes!

	 On aurait tort de penser que cette valse hésitation est réservée à
	 Einstein ou a Bohr. Voici en quelque sorte ``l'Institution''
	 maintenant.
	 
	 {\bf 1913} Dans une lettre de recommandation de Planck, Nerst, Rubens et
	 Warburg pour soutenir la candidature d'Einstein à l'académie des
	 sciences de Berlin :

	 \bq ``En bref, on peut dire, parmi les grands problèmes dont la
	 physique moderne abonde, il n'en est guère qu'Einstein n'ait marqué
	 de sa contribution. Il est vrai {\bf qu'il a parfois manqué le but
	 lors de ses spéculations, par exemple avec son hypothèse des
	 quantas lumineux}. Mais on ne saurait lui en faire le reproche, car
	 il n'est pas possible d'introduire des idées réellement nouvelles,
	 même dans les sciemces les plus exactes, sans prendre parfois des
	 risques''{ \em\cite{bali1989} page 38}.\eq

On est cette fois huit ans après la fondation et les quanta sont encore pour
l'Institution une spéculation farfelue!

	{\bf 1916} Onze ans maintenant, et on va trouver là une préoccupation qui
	durera toute la vie d'Einstein (voir plus loin une lettre de
	1953-54) la recherche d'une possibilité de contourner les
	difficultés de la mécanique quantique en en changeant les bases et
	en particulier pourrait-on dire la tentation récurrente pour lui du
	discontinu :

	\bq ``Comment formuler des énoncés relatifs au discontinu sans avoir
	recours à un continuum -l'espace-temps- ; ce dernier devrait être
	exclu de la théorie, en tant qu'il est une construction adventice
	que ne justifie pas l'essence du problème et qui ne correspond à
	rien de réel. A cet égard, nous manquons cruellement de formalisme
	mathématique adéquat'' {\em Lettre à Dällenbach \cite {bali1989} page
	232}.\eq

Ce n'est pas un mince changement ; rien que l'abandon de l'espace-temps. Et
si Planck avait eu connaissance de cette spéculation supplémentaire!?

	En {\bf 1922} Bohr conteste toujours l'aspect corpusculaire du rayonnement
	et propose dans un article de {\bf 1924} avec Kramers et Slater de
	renoncer à la conservation de l'énergie et de l'impulsion dans
	chaque collision mais de ne la respecter seulement qu'en moyenne.
	Une spéculation de plus, mais celle de Bohr cette fois!

	Einstein qui avait envisagé et abandonné cela en 1910 (12 ans
	auparavant, voir plus haut) conteste évidemment et, compte tenu de la
	notoriété de chacun des participants, la polémique atteint le grand
	public par la presse.

	C'est en {\bf 1925} que l'expérience de Bothe et Geiger, avec la
	diffusion Compton, tranchera en faveur d'Einstein : les lois de
	conservation de l'impulsion et de l'énergie sont respectées à
	l'échelle des collisions individuelles.

        Pourtant ce n'est que deux ans plus tard, en {\bf 1927}, qu'on
	trouvera cette déclaration péremptoire de Born et Heisenberg,
	rappelons la :

\bq"Nous tenons la mécanique des quanta pour une théorie complète dont
les hypothèses fondamentales physiques et mathématiques ne sont plus
susceptibles de modifications."\eq

\section {	Einstein et la mécanique quantique : quatre points forts.}

	1) La description que fournit la mécanique quantique et les
	prédictions qui s'en déduisent sont justes.

	2) La mécanique quantique est insuffisante, inachevée, incomplète
	car elle ne traite pas les cas individuels.

	3) Pas de bricolage possible, pas de paramètres supplémentaires.
	Pour arriver à la théorie complète il faut une refondation, une
	reconstruction.

	4) Démonstration de l'alternative EPR et la question des actions à
	distance.

On peut dire qu'on trouve la trace des trois premiers points tout au long du
demi-siècle qui sépare 1905 de 1955 ; pour le quatrième, c'est plus compliqué.
S'il est forcément schématique de rigidifier ainsi en quatre points ce que
furent cinquante ans d'interventions sur le sujet, au moins, ce schéma
est-il construit à partir de textes qu'on peut vérifier!
Examinons donc successivement ces quatre points.

\subsection *{1) La mécanique quantique dit juste :}

	\bq " Cette théorie est jusqu'à maintenant la seule qui unifie le
	double caractère corpusculaire et ondulatoire de la matière d'une
	façon dont la logique est satisfaisante ; et les relations
	(vérifiables) qu'elle contient, sont, à l'intérieur des limites
	fixées par la relation d'incertitude {\em complètes}. Les relations
	formelles qui sont données dans cette théorie -c.a.d. son formalisme
	mathématique tout entier- devront probablement être contenues, sous
	la forme de conséquences logiques, dans toute théorie future
	utile"{\em\cite{eins1949} Einstein's reply,  page 666-667. }\eq

C'est une approbation profonde qui ne devrait pas permettre en 2005 le titre
de chapitre de ``Pour la Science' cité plus haut :

{\Large  "Einstein n'accepte pas la mécanique quantique}"

\subsection *{ 2 La mécanique quantique dit juste, mais elle ne dit pas tout.}

	\bq "Je suis en fait, et au contraire de presque tous les physiciens
	théoriciens contemporains fermement convaincu que le caractère
	essentiellement statistique de la théorie quantique contemporaine
	doit uniquement être attribué au fait que cette théorie opère avec
	une description incomplète des systèmes
	physiques''{\em\cite{eins1949}Einstein's reply, page 666. } \eq
	
	ou encore
	
	\bq ``Ce qui ne me satisfait pas dans cette théorie, du point de vue du
	principe, c'est son attitude envers ce qui m'apparaît comme le but
	programmatique de toute la physique : la description complète de
	toute situation réelle (individuelle) comme il est supposé qu'elle
	existe indépendamment de tout acte
	d'observation"{\em\cite{eins1949}Einstein's reply, page 667. } \eq

Là, c'est vrai, c'est autant l'exigence du caractère objectif des phénomènes
physiques qui est rappelée à côté de celle des situations individuelles.
Mais on va retrouver maintenant, en 1953, le "dieu ne joue pas aux dés" de
1926 avec une expression plus rigoureuse ... et moins humoristique :

	\bq "En fin de compte, l'idée que la physique doit s'efforcer de
	donner une "description réelle" d'un système individuel est bien
	incontournable. La nature, prise comme un tout, ne peut être pensée
	que comme un système individuel (existant de façon unique) et non
	comme un {\em ensemble de systèmes"  Hommage à Max Born,
	\cite{bali1989} page 256. } \eq

\subsection *{ 3) Pas de bricolage, un aménagement ne peut suffire.}

	\bq " Je ne pense pas que l'on puisse arriver à une description des
	systèmes individuels simplement en complétant la théorie quantique
	actuelle. Le principe de superposition et l'interprétation
	statistique sont indissociablement liés entre eux. Si l'on pense que
	l'interprétation statistique doit être dépassée, on ne peut pas
	conserver l'équation de Schrödinger dont la linéarité implique la
	{\em superposition des états" Lettre à Kupperman de novembre 1953, \cite{bali1989}
	page 233.} \eq
	
On peut déduire de cela que les théories à variables supplémentaires ne
conviennent pas à Einstein ni même celle de Bohm qui pourtant est causale et
traite donc comme souhaité des {\em cas individuels}. Écrire comme Bell, dans
son article fondateur sur les inégalités :
\bq ``Le paradoxe EPR a été proposé comme un argument comme quoi la mécanique
quantique ne pouvait pas être une théorie complète mais devait être
complétée par des variables supplémentaires, ces variables devant restaurer
la causalité et la localité''{\em \cite{bell1964}}\eq
ne semble donc pas conforme à la réalité de la pensée d'Einstein, au moins,
telle qu'elle est exprimée en 1953. Bell précise un peu plus loin :
\bq
``C'est l'exigence de localité, ou plus précisément que le résultat d'une
mesure sur un système ne soit pas affecté par des opérations sur un système
distant avec lequel il a interagi dans le passé, qui crée la difficulté
essentielle''.
\eq

C'est vrai, comme on le verra plus loin, mais cela n'a rien à voir avec
l'exigence du traitement des cas individuels comme l'exige Einstein.
Rappelons, avec cette fois la formulation de 1954 (on a vu 1916!) le genre
de refondation qu'Einstein peut penser nécessaire ou au moins ne craint pas :

	\bq "Il me semble en tout cas, que l'alternative continu-discontinu est
	une authentique alternative ; cela veut dire qu'ici, il n'y a pas de
	compromis possible. ... Dans cette théorie, il n'y a pas de place
	pour l'espace et le temps, mais uniquement pour des nombres, des
	constructions numériques et des règles pour les former sur la base
	de règles algébriques excluant le processus limite. Quant à savoir
	quelle voie s'avérera la bonne, seule la qualité du résultat nous
	l'apprendra"{\em Lettre à Joachim, \cite{bali1989} page 256.} \eq
	
On trouvera par exemple dans\cite{avra2003} une recherche qui s'inscrit dans
cette perspective.
On est en effet très loin des variables supplémentaires!

\subsection *{ 4) L'alternative EPR}

On va voir dans le prochain chapitre la démonstration de l'alternative EPR
et ses conséquences. Mais cette fois, clairement, la position d'Einstein
n'est pas la même en 1935 et en 1949 ; elle a peut-être même varié entre
1946 et 1949!

\section {4) Einstein, Podolsky, Rosen. Démonstration de l'alternative EPR
et la question des actions à distance.}

On va commenter l'expérience de pensée proposée par Einstein, Podolsky et
Rosen et schématisée sur les deux pages suivantes.

\newpage
\Large

$~$
\vspace{-2.5cm}

{\LARGE \em L'expérience de pensée EPR}

\vspace{1cm}

\hspace{5cm}
\begin {minipage}{10cm}
La relation de Heisenberg $\Delta x \Delta v \geq$ Cte

\begin{pspicture}(.4,.2)
\psline[linewidth=.2]{->}(0,0.25)(0.5,.25)
\end{pspicture}
{\bf Des états limites sont possibles} :

{\bf * `états x'} : x prend une valeur précise, v est indéterminé

{\bf * `états v'} :  v prend une valeur précise, x est indéterminé
\end{minipage}

$~$

\vspace{4cm}

{\bf Première série d'expériences :\\
\hspace{4cm} mesures de {\em position} sur B.}

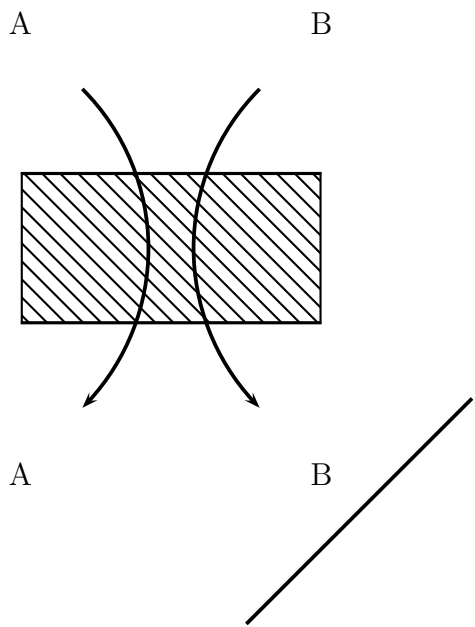
\vspace{4cm}
\begin{pspicture}(5,5)
\rput(0,1){\large A}
\rput(0,7){\large A}
\rput(4,1){\large B}
\rput(4,7){\large B}
\psline{-}(0,3)(4,3)
\psline{-}(0,5)(4,5)
\psline[linewidth=.05]{-}(3,-1)(6,2)
\psframe[fillstyle=vlines] (0,3)(4,5)
\rput(2.,4.){
\psarc[linewidth=.05]{<-}(-3.3,0){3}{-45}{45}
\psarc[linewidth=.05]{->}(3.3,0){3}{135}{225}
}
\rput(13,0){
\begin {minipage}{10cm}
On mesure la position de B ;\\ on trouve par exemple x$_{B}$, mais\ldots  

A est {\bf aussi} dans un ``état x''

(x$_{A}$ = x$_{B}$ + x$_{0}$, par exemple)
\end  {minipage}
}
\end{pspicture}

\newpage 

$~$

\vspace{4cm}

{\bf Deuxième série d'expériences ;\\
\hspace{4cm}mesures de {\em vitesse} sur B.}


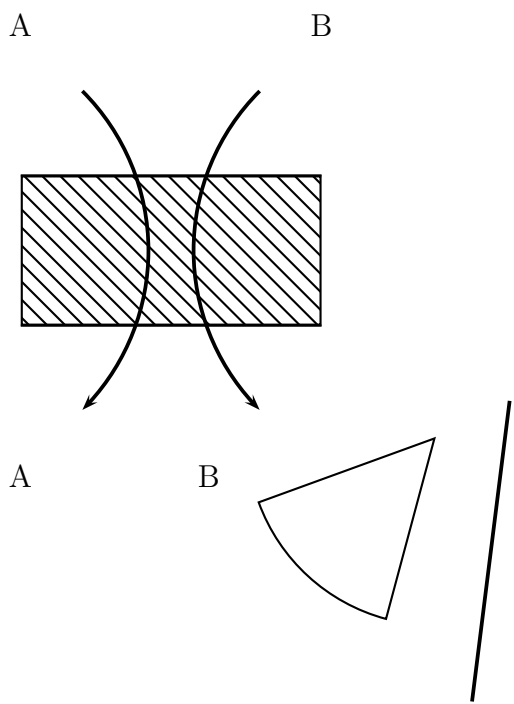
\begin{pspicture}(11,11)
\rput(0,1){\large A}
\rput(0,7){\large A}
\rput(2.5,1){\large B}
\rput(4,7){\large B}
\psline{-}(0,3)(4,3)
\psline{-}(0,5)(4,5)
\psline[linewidth=.05]{-}(6,-2)(6.5,2)
\psframe[fillstyle=vlines] (0,3)(4,5)
\rput(2.,4.){
\psarc[linewidth=.05]{<-}(-3.3,0){3}{-45}{45}
\psarc[linewidth=.05]{->}(3.3,0){3}{135}{225}
}
\pswedge(5.5,1.5){2.5}{200}{255}

\rput(13,0){
\begin {minipage}{10cm}
On mesure la vitesse de B ;\\
on trouve par exemple v$_{B}$, mais\ldots  

A est {\bf aussi} dans un ``état v''

(v$_{A}$ = -v$_{B}$, par exemple)
\end  {minipage}
}

\end{pspicture}
\normalsize

\newpage 

A cause, ou grâce au principe d'incertitude de Heisenberg, on ne peut
définir à la fois la position et la vitesse d'une particule :

$\Delta x  \Delta v \geq \hbar/m = Cte$            (Heisenberg)

On peut cependant définir deux cas  limites, deux états limites :

	des "{\bf états x}" pour lesquels x est très bien connu ($\delta$ x est
	petit) et v est complètement indéterminé

	des "{\bf états v}" pour lesquels v est très bien connu ($\delta$ v est
	petit) et x est complètement indéterminé

et le principe d'Heisenberg dit justement qu'une particule ne peut être à la
fois dans un "état x" et dans un "état v".

Préparation (autant de fois qu'on le souhaite, de façon identique) A et B
sont entièrement connus et décrits par une fonction d'onde.

On laisse interagir A et B pendant un certain temps et puis l'interaction
est interrompue (ce qui pourrait se passer au cours d'une collision par
exemple).

Après cette interaction, ni A ni B n'ont de fonction d'onde car c'est le
système composite qui existe et est décrit par une fonction d'onde globale,
qui ne se factorise pas. Pour retrouver une fonction d'onde pour chacune des
composantes A et B, il faut effectuer une "mesure" sur l'une ou sur l'autre
des composantes (A ou B) ; une "réduction du paquet d'onde" doit se produire.
Cette mesure cependant a des conséquences observables non seulement sur le
composante B par exemple) sur lequel elle a été effectuée, mais aussi sur
l'autre (A dans l'exemple). Et EPR précisent encore l'expérience de pensée
en notant que deux types de mesure sont possible, mesure de x ou mesure de v.

Le premier type de mesure conduit à un ``état x'' pour B et,à cause de la préparation et
c'est le coeur de l'argument EPR, répétons le, {\bf aussi} à un ``état x''
pour A, par exemple xA=xB+x0.

Le second conduit à un ``état v'' pour B et (de la même façon que précédemment
pour x) à un ``état v'' pour A, par exemple, vA=-vB.

d'où l'alternative EPR :

$~$

\fbox{\parbox{15cm}{
1)ou bien xA et vA ont tous les deux une valeur déterminée dans le passé
commun (avant de se séparer, à la fin de leur interaction)

2) ou bien il y a "interaction après l'interaction'' : l'état de A dépend de
l'état de B ; A "communique" avec B par "télépathie" (on reviendra plus loin
sur ce terme!)}}

1) contredit la mécanique quantique mais pas ses prédictions : la mécanique
quantique est incomplète, elle n'est pas fausse. Insistons que l'aspect
statistique de la mécanique quantique n'est alors en rien affecté par ce
choix dans l'alternative. Ce premier terme si on le choisit, conduit
effectivement à compléter la mécanique quantique, mais il n'a pas pour {\em
résultat} de supprimer l'indéterminisme ; il a pour résultat de renvoyer cet
indéterminisme {\em dans le passé commun} des deux particules. En
démontrant, en pensant démontrer, que, à un moment donné (après leur
séparation) chacune des particules "possédait" {\bf et} une vitesse {\bf et}
une position, il pulvérisait les bases de la mécanique quantique mais sans
en changer les résultats ni les prédictions, ni donc la conformité à
l'expérience. C'est en fait la cohérence qu'il mettait en défaut même si ce
n'est pas ce mot qu'Einstein a alors employé.

2) contredit... le bon sens\ldots   et/ou complète radicalement la physique.

Il pose de nouveau et autrement la question des actions à distance, de la
``non-localité' apparues très tôt des l'établissement de la dualité
onde/corpuscule ou dans l'examen des expériences de diffraction, diffraction
à une ou à deux fentes.

\section {Quelles réponses à l'alternative EPR.  }

Personne n'est bien sûr au dessus de la mêlée, au mieux peut-on espérer
être dedans! Sur les questions soulevées par EPR on va donc supposer, en ce
qui concerne la physique, que c'est ce qui se dit aujourd'hui qui est le
plus proche de la réalité, de la vérité, jusqu'à ce qu'un autre progrès
vienne présenter demain les choses autrement.

\subsection *{La référence : ce qu'on dit aujourd'hui.}
Et donc aujourd'hui et avec Alain Aspect par exemple, on va parler, à partir
des expériences et de l'interprétation qu'on en donne aujourd'hui d' {\bf
\Large "états intriqués de particules distantes"} et de {\bf \Large
''corrélations fortes"} de leurs propriétés ou encore que deux particules
qui ont interagi forment un {\bf \Large ``système unique''}.

Confrontons alors ce point de vue actuel avec successivement les
participants du débat et d'abord bien sûr avec Einstein qui en est à la
source, et d'abord en 1935 avec l'article EPR lui même :

\subsection *{Les termes d'Einstein. En 1935, puis en 1946 et 1949.}
	\bq "Cela fait que la réalité [état x ou état v] du système A dépend
	d'un processus de mesure effectué sur le système B avec lequel le
	système A n'interagit pas. Aucune définition de la réalité ne
	pourrait permettre cela"{\em \cite{EPR} page 780.}\eq

Einstein met bien en contradiction le "dépend de" avec "n'interagit pas".
Notons qu'il n'est question que de physique. Et  dans les termes de 1946, il
précise la seule échappatoire :

	\bq "La mesure sur B change par télépathie la situation réelle de
	A"{\em \cite{eins1949}Einstein's autobiographical notes, page 85.}\eq

Et jusqu'en 1949, clairement, il refuse cette télépathie, cette "action à
distance" dont on peut dire qu'elle est la préfiguration de ces états
intriqués, de ces corrélations fortes à distance.  Mais voilà  comment en 1949
Einstein termine toute la partie de son intervention consacrée à la
mécanique quantique :

	\bq " Je vais terminer ces présentations qui se sont développées plutôt
	longuement, concernant l'interprétation de la mécanique quantique
	par la reproduction d'une brève conversation que j'ai eue avec un
	physicien théoricien important :
	
	Lui :  ` J'ai tendance à croire à la télépathie.'
	
	Moi : `Ceci a probablement plus à voir avec la physique qu'avec la
	psychologie.'
	
	Lui : `Oui'.
	
	{\em \cite{eins1949}Einstein's reply, page 683.}\eq

La place donnée à ce qui pourrait n'être qu'une boutade oblige à penser
qu'en 1949, au moment de l'écriture de ce document, Einstein doute. Par
ailleurs, on peut penser que c'est avec David Bohm qu'a été tenu cet
échange. Mais une fois de plus, c'est bien de physique dont il s'agit, et
pas de philosophie.

\subsection *{ La réponse de Schrödinger en 1935.}

Parce qu'elles ont été discutées par ailleurs \cite{rous2002}, on ne
reparlera pas ici des positions de Schrödinger pourtant très importantes ;
c'est lui qui a introduit la notion d'intrication (Verschränkung,
entanglement, enchevêtrement... intrication) en parfaite résonance avec la
physique d'aujourd'hui. Mais, dans la plus grande clarté, même si c'est à son
corps défendant, il place le problème du côté de l'abstraction : la fonction
d'onde est un catalogue des résultats possible de mesure et l'intrication
est celle de ses résultats. Ainsi, la question des actions à distance ne se
pose pas, elle est sortie de la physique.

\subsection *{La réponse de Bohr en 1935\ldots  confirmée en 1946. }

\bq" Mon objectif principal, ... , c'est d'insister sur le fait que nous
n'avons pas à faire à une description incomplète caractérisée par le choix
arbitraire de différents éléments de la réalité physique au sacrifice
d'autres tels éléments mais par la discrimination rationnelle entre des
systèmes expérimentaux et des procédures différentes qui sont adaptées, soit
à l'utilisation non ambiguë de la notion de localisation soit à l'utilisation
légitime du théorème de conservation de l'impulsion. Tout ce qui reste qui
puisse paraître arbitraire concerne simplement {\bf \large notre liberté de
manoeuvrer les instruments de mesure}, une caractéristique de l'idée même
d'expérience.

En fait, le renoncement dans chaque arrangement expérimental, de l'un ou
l'autre des deux aspects de la description des phénomènes physiques, -dont
la combinaison caractérise la méthode de la physique classique et qui donc
dans ce sens peut être considérée comme complémentaire l'un de l'autre-
repose essentiellement sur l'impossibilité, dans le champ de la théorie
quantique, de {\bf \large contrôler précisément la réaction de l'objet sur
l'instrument de mesure}, i.e. le transfert de moment dans le cas d'une
mesure de position et le déplacement dans le cas d'une mesure de moment...
En vérité, nous avons dans chaque arrangement expérimental adapté à l'étude
d'un phénomène proprement quantique, non pas simplement à faire avec
l'ignorance de la valeur de certaines quantités physiques, mais avec
l'impossibilité de définir ces quantités de façon non ambiguë''{\em
\cite{bohr1935} page 699.

et un peu plus loin encore :}

``A partir de notre point de vue, nous voyons maintenant que les mots même du
critère de réalité physique proposé par EPR contient une ambiguïté en ce qui
concerne le sens de l'expression " sans du tout perturber un système". Bien
sûr, il n'y est dans un cas comme celui là, pas question d'une perturbation
mécanique du système en investigation durant la dernière étape de la
procédure de mesure. Mais même à cette étape, c'est la question d'{\bf
\large une influence des conditions mêmes} qui définissent les types
possibles de prédictions concernant le comportement futur du système''{\em
\cite{bohr1935} page 700 }.\eq

Peut-être alors est-on tenté aujourd'hui de reprendre à notre compte ce
jugement plutôt sévère d'Einstein :

\bq ``C'est pour moi une faute de permettre à des descriptions théoriques d'être
directement dépendantes d'assertions empiriques comme dans le principe de
complémentarité de Bohr à la formulation précise duquel je ne suis pas
parvenu malgré beaucoup d'efforts que j'y ai consacré''{\em \cite
{eins1949}Einstein's reply, page 674 .}\eq

Si les citations de N. Bohr ci-dessus sont toutes extraites de sa réponse à
EPR en 1935, elles ont été confirmées en 1946 quand N. Bohr a repris
(volontairement!) les mêmes arguments \cite{eins1949} Discussion with
Einstein on epistemological problems in atomic physics ,pages 201-241.

\subsection *{ Ce que disent D. Bohm et A. Aharonov. }

Bohm et Aharonov s'appuyant sur des résultats expérimentaux et théoriques
antérieurs (1946, 1950 et 1951) montrent, avec les photons gammas (511 keV)
issus de la désintégration des électrons positifs, que l'état des deux
photons n'est pas ``factorisé'' et qu'ils se comportent donc selon les
prédictions annoncées par EPR et \ldots mises en doute par EPR!

	\bq "Une brève revue est donnée de la signification physique du
	paradoxe ERP (sic) et il est montré que cela implique un type de
	corrélation des propriétés de systèmes distants non interagissant,
	qui est complètement différent des types de corrélations connus.
	
	\ldots 
	
	Ainsi cette expérience {\em [celle qui est rapportée et interprétée dans
	cet article]} peut être regardée comme la première preuve empirique
	claire que les aspects de la théorie quantique décrits par ERP (sic)
	représentent des propriétés réelles de la matière"{\em
	\cite{bohm1957} resumé;}

{\em et plus loin :}

	" On pourrait peut-être supposer qu'il y a une interaction cachée
	entre B et A, ou entre B et l'appareil de mesure [sur A?], qui
	explique le comportement observé. Une telle interaction serait pour
	le moins en dehors du domaine de la théorie quantique courante. De
	plus, elle devrait être instantanée,{\bf parce que les orientations
	de l'appareil de mesure pourraient être changées très rapidement, et
	le spin de B devrait répondre immédiatement à ce changement}. Une
	telle interaction immédiate entre des systèmes distants ne serait
	pas en général compatible avec la théorie de la relativité. Ce
	résultat constitue l'essence du paradoxe ERP"{\em \cite{bohm1957}
	page 1071.}\eq

Avec Bohm, on est tout à fait en résonance avec la problématique actuelle ;
elle y est même exprimée avec la plus grande clarté.

\subsection *{ En résumé donc :}

\begin{itemize}
\item A. Einstein.  Met en évidence la physique de l'intrication mais en
refuse d'abord les conséquences, la "télépathie", jusqu'en 1946... puis
exprime un doute à partir de 1949.

\item E. schrodinger. La fonction d'onde comme "catalogue" ; ``intrication''
oui, mais des résultats de mesure ; on reste dans l'abstraction ; pas de
``télépathie''.

\item N.Bohr   La complementarité (des appareils de mesure) ; l'action
incontrôlable (des appareils de mesure) mais pas de "télépathie", pas même
intrication?
\item D. Bohm La physique et l'expérience qui lève le doute.
\end{itemize}

Ne doit-on pas dire alors que Bohr parce qu'il ne répond à EPR que par
les deux concepts de complémentarité et de transfert incontrôlable refuse
l'alternative EPR plus qu'il n'en choisit le bon terme. Il ne peut donc pas
être considéré comme à l'origine de cette part de physique qu'on décrit
aujourd'hui avec les états intriqués.

 On peut également dire que Aspect, Einstein, et Bohm ont ceci de commun
 qu'ils restent dans la physique alors que Schrodinger et Bohr chacun à sa
 façon s'en échappent ; mais Schrodinger a reconnu la nouveauté de
 l'intrication alors que Bohr la fuit, plus peut-être qu'il ne la refuse.

Dire alors que la confirmation des corrélations EPR a donné raison à Bohr
contre Einstein comme il est répété à foison ne parait pas conforme aux
faits. Plus certain encore c'est que cet aspect de la physique quantique est
totalement indépendant de la question des ``cas individuels'' c.a.d. de
l'aspect statistique de la mécanique quantique.

\section { Conclusion }
Reprenons ce que disait Françoise Balibar déjà citée en introduction :

	\bq ``Einstein est l'auteur de plusieurs théories qui n'ont jamais
	été démenties. C'est même extraordinaire qu'au bout de cent ans ces
	théories n'aient pas été supplantées par d'autres, plus actuelles.
	Einstein est donc d'actualité parcequ'il a mis en route une théorie
	dont notre univers dépend. Mais aussi parcequ'en énonçant des
	objections à ce que devenaient ses travaux, il a semé les germes de
	nouveaux développements de ses théories. Dans la première moitié du
	XXième siècle, on considérait pourtant que ses objections étaient
	celles d'un vieil homme décalé par rapport à son
	temps''{\em\cite{huma2005}}.\eq

Les états intriqués et tous les développements théoriques et expérimentaux
qu'ils ont suscités illustrent parfaitement ce propos.

Je me suis efforcé de donner quelques repères forcément bien partiels mais
que j'espère représentatifs sur les rapports critiques d'Einstein à la
mécanique quantique et de ses rapports sur ce sujet avec ses contemporains.
J'ai essayé d'éclairer leur contenu en examinant leur rapport avec la
physique actuelle. Je voudrais encourager ceux que le sujet intéresse à
aller eux même consulter les sources : elles sont pleines de richesse et de
surprises.

Je voudrais terminer par trois points qui sont, ou sont encore, ou devraient
être l'objet de discussions :

\begin {enumerate}

\item Il semble bien, comme suggéré dans le résumé, que si le jugement porté
sur Einstein et la mécanique quantique change -rappelez vous Heisenberg et
Aspect- c'est qu'en effet le débat évacué vers le philosophique redescend ou
remonte, disons se recentre sur la physique. Il ne semble pas que ce
mouvement soit terminé comme peuvent en témoigner les quelques citations
plus ou moins contemporaines qui suivent.

\bq
Selon nos calculs, on peut décider de mettre en évidence soit le
comportement ondulatoire (interférence), soit le comportement de particule
(quelle trajectoire) même après le temps de l'émission,{\bf sans manipuler
physiquement les photons  $\gamma$}
\cite{Scul82} \eq

\bq
Pourtant il {\em [le photon]} se comporte toujours d'une façon qui dépend du
test effectué sur l'autre photon, bien {\bf qu'il ne puisse être
physiquement influencé} par l'accomplissement de la mesure ou par le
résultat ainsi obtenu {\em \cite{Ghir02}.}
\eq

\bq
On reviendra sur cette question\ldots pour décrire comment l'expérience
[EPR] a été réalisée en soulignant ce qui semble faire aujourd'hui son
intérêt essentiel : la mise en évidence d'états ``enchevêtrés'' à longue
distance \ldots avec toutes leurs {\bf diableries} surprenantes mais
avérées {\em \cite{Omne00}.}
\eq

\bq
Cette intrication subsiste même si les deux atomes se sont éloignés l'un de
l'autre et se trouvent séparés après la collision par une distance
arbitrairement grande. Elle décrit une ``non-localité'' fondamentale de la
physique quantique. Une mesure de l'atome 1 peut avoir un effet immédiat à
grande distance sur le résultat de la mesure de l'atome 2! Il y a donc entre
les deux particules un lien quantique {\bf immatériel et instantané}. C'est
Einstein, avec ses collaborateurs Podolsky et Rosen, qui a discuté le
premier en 1935 cet aspect troublant de la théorie quantique {\em
\cite{Haro01} page 577.}
\eq

Mais justement, peut-être Einstein répondrait-il à tous ces physiciens que
c'est de physique qu'il s'agit et d'une part que l'immatériel n'a pas sa
place et que d'autre part, l'instantané pose problème.

\item  Sur la question de l'indéterminisme quantique, on trouve de façon
répétée ce jugement exprimé ici dans ce livre déjà cité :

	\bq ``En mécanique quantique, seules ces probabilités nous sont
	accessibles : il n'y a pas de déterminisme caché sous une apparence
	statistique, ce qui a été ensuite démontré expérimentalement par les
	travaux d'Alain Aspect''{\em\cite{lcp2005}.}\eq

Ceci n'est pas exact et probablement ne peut pas l'être car on ne démontre
pas le hasard! Mais revenons aux quatre point forts de la position
d'Einstein : le point 4) (EPR) visait à démontrer le point 2), pas à lui
apporter une solution. La solution (à rechercher selon Einstein) c'est le
point 3), la refondation. La démonstration 4) a dévié ; d'un côté elle a
conduit à mettre en évidence des propriétés nouvelles de la matière comme
le disait Bohm, d'un autre, elle a en quelque sorte échoué à démontrer
autrement le point 2). Cet échec n'invalide absolument pas le point 2) et
encore moins en démontre le contraire (une démonstration qui échoue ne
prouve rien que son échec).

Mais d'une certaine manière, l'unicité du monde qui fonde pour Einstein la
nécessité de traiter "les cas individuels" (et pas seulement les
ensembles) est établie par la physique mise en évidence par le second
terme de l'alternative EPR. Celle-ci en effet, au travers de
"l'intrication", rend de proche en proche, dans le temps et dans
l'espace, toute particule dépendante de toutes les autres.

\item Enfin, ~on peut rappeler pour terminer quels sont les problèmes de la
mécanique quantique qui restent posés aujourd'hui en 2005, même s'ils le
sont autrement qu'en 1905 :
\begin{itemize}

\item  la mesure, le statut de la fonction d'onde, les cas individuels, la
refondation?(le discontinu?).

\end {itemize}
\end {enumerate}

Le nouveau cependant, c'est bien comme le déclarent les 
intervenants de l'ENS :
\bq
``On peut alors aborder à nouveau, mais de façon concrète, l'étude des
fondements de la théorie''{\em Voir par exemple \cite{Haro01} page 572.}
\eq
Il n'y a pas de doute que Einstein s'en réjouirait mais qu'il ne manquerait
pas non plus de rappeler :

\bq
Une théorie peut être mise à l'épreuve par l'expérience, mais il n'y a pas
de chemin qui conduit de l'expérience à l'établissement d'une
théorie''{\em \cite{eins1949} Autobiographical notes, page 88-89.}
\eq

En mars 1979, lors de la célébration du centenaire de sa naissance, Banesh
Hoffmann écrivait dans ``Physics today'' (page 40) ;

\bq
``Il fut un temps où c'était pratiquement un suicide professionnel pour un
physicien de manifester des doutes à propos de l'interprétation dénommée de
Copenhague. Depuis lors, le climat est moins tendu même si la plupart des
physiciens acceptent l'interprétation de Copenhague et croient qu'il n'y
aura jamais de retour au déterminisme ancien\ldots Dans la perspective de la
centaines d'années qui nous séparent du bicentenaire, la défiance intuitive
d'Einstein pour les idées quantiques probabilistes courantes pourraient
apparaître se justifier dans un sens totalement inattendu \ldots  comme se
fut le cas pour la proposition des quantas de lumière en 1905 décriée
pendant si longtemps.''
\eq

Seulement 25 ans nous séparent de cette déclaration!

\end{document}